\documentclass[11pt,onecolumn,amssymb,nofootinbib]{revtex4}
\usepackage{amsmath, amsthm, amscd, amssymb}
\usepackage{graphicx, braket}
\usepackage{bm}
\usepackage{bbm}
\usepackage{epstopdf}

\begin{document}

\title{\bf Black hole mimickers as relativistic stars calculated from the Tolman-Oppenheimer-Volkoff equations}

\author{Stephen L. Adler}
\email{adler@ias.edu} \affiliation{Institute for Advanced Study,
Einstein Drive, Princeton, NJ 08540, USA.}

\begin{abstract}
This mini-review summarizes and simplifies the principle content of our papers on ``Dynamical Gravastars'',  formulated using the Tolman-Oppenheimer-Volkoff equations,  and surveyed numerically using Mathematica notebooks that we document here in Appendices.
\end{abstract}

\maketitle
\section{Introduction and Historical Remarks }

It now seems likely that each of the trillions of galaxies in the visible universe is centered on a supermassive astronomical ``black hole'', and counting smaller black holes there may be as many as $4 \times 10^{19}$ black holes present.
 But the key question remains of whether these are true mathematical black holes, each bounded by a one-way horizon, or whether these are some type of ``black hole mimicker'' without a horizon.  There is already a large literature on black hole mimickers, for which we refer the reader to the review articles \cite{pani}, \cite{mottola1}.  Our aim in this mini-review is  to sketch the principal results on the genre of black hole mimickers that we have termed ``dynamical gravastars'', which appeared in three papers that we have written \cite{adler1}, \cite{adler2} or co-written \cite{adler3}, and in a memorandum on our home page \cite{adler4}.

The first of these papers \cite{adler1} can be confusing to read, because in it we incorporated the suggestion of a horizonless black hole arising from a scale invariant cosmological constant action \cite{rev1} explored in an earlier paper by Adler and  Ramazano\u glu \cite{fethi} in the absence of matter.  This was combined in \cite{adler1} with an investigation of the solutions of the Tolman-Oppenheimer-Volkoff (TOV) equations with matter undergoing an  equation of state jump.  The TOV investigation led to an unexpected  second mechanism for generating a horizonless black hole, which is still operative and  simpler when the cosmological constant is set to zero, as was done in our subsequent papers \cite{adler2}, \cite{adler3}, and \cite{adler4}.  In this mini-review we limit ourselves to a discussion of the simplified model with no cosmological constant.

The idea that black hole interiors may result from an interior phase transition is an old one, going back at least to the suggestion by Gliner \cite{gliner} of a phase transition  from a normal matter equation of state to a ``gravity vacuum'' equation of state, in which the pressure $p$ is minus the density $\rho$. This is the vacuum equation of state associated with a pure cosmological constant or de Sitter universe, and gives rise to the name {\bf gravastar} = {\bf gra}vity {\bf va}cuum {\bf star}.  Related ideas were discussed via a condensed matter analogy in \cite{other}, \cite{other2}, \cite{khlopov1}, \cite{khlopov2}.  The idea of gravastars has gained particular prominence from the seminal  papers of Mazur and Mattola \cite{mazur}, which are based on assuming a pressure jump in the interior equation of state, from a normal matter equation of state to the ``gravity vacuum'' equation of state proposed by Gliner \cite{gliner}.  In our papers we have adopted the Gliner suggestion (which as we shall see has an alternative non-cosmological motivation coming from the TOV equations) but we believe that the idea of a pressure jump is incorrect; in systems in which all physical quantities are bounded the pressure must be continuous.  Instead, we exploit the fact that when quantum effects are included, the energy density $\rho$  can take values smaller than the pressure $p$ \cite{ruderman}, and even negative values \cite{wald}, permitting an energy density jump and thus providing the foundation of our calculations.

\section{Energy Conditions and the Gliner vacuum}

Four classical energy conditions are introduced in discussions of general relativity.  In terms of the pressure $p$ and energy density $\rho$ of an isotropic fluid\footnote{For the general non-isotropic case, see Sec. 4.3 of Hawking and Ellis \cite{hawk}.}  they are:

\begin{itemize}

\item The ``weak energy condition''  $\rho \geq 0,\, \rho+p \geq 0$,

\item The ``dominant energy condition''  $\rho \geq |p|$,

\item The ``null energy condition''  $\rho+p \geq 0$,

\item The ``strong energy condition''  $\rho+p\geq 0,\, \rho+3p \geq 0$.

\end{itemize}

The Gliner vacuum condition $\rho+p= 0$ cannot simultaneously satisfy all four of the classical energy conditions.   The dominant energy condition requires $\rho$ to be nonnegative, and in fact positive if the pressure is nonzero. But in that case the Gliner condition requires $p=-\rho <0$, and the second part of the strong energy condition becomes $\rho+3p=2p <0$, violating the strong energy condition.   So the Gliner vacuum cannot satisfy both the dominant energy condition and the strong energy condtion.  One has to choose which of the two is to be retained in a theory of black hole mimickers.

The choice in the papers of  Mazur and Mottola and many others that follow their analysis is to impose the dominant energy condition, with the pressure making a transition  from a normal matter positive value to a negative value in the mimicker interior, with the transition taking place over  thin ``skin'' layers.  The radii characterizing these transition layers are then model parameters, which seem to us to be an unnecessary complication.   Moreover, in the limit of vanishing thickness of a skin layer, so that the pressure has a discontinuous jump, the energy density associated with that skin layer must become infinite to avoid acceleration of the skin by the pressure differential on the two sides of the layer.

The choice in our papers is to take the strong energy condition, as well as the weaker null energy condition,  as the ones to be preserved.  The Gliner vacuum then is attained by having the pressure remain always positive, while the energy density in the mimicker interior becomes negative.  It has long been known that when quantum corrections to the energy density are included, with consequent renormalizations, the classical energy no longer needs to be positive \cite{wald}, so the assumption of negative interior energy density is physically allowed.  This of course gets away from the Gliner motivation of a cosmological constant vacuum as a model for the interior, but we will see in our review of the TOV equations that the condition $p+\rho=0$ has a natural interpretation as the limit of vanishing pressure evolution in these equations. We also note that the Penrose and Hawking singularity theorems are based  on the null  and/or  strong energy conditions, with the dominant energy condition not used in their derivations.  So sacrificing the dominant energy condition seems to us a reasonable choice.\footnote{As emphasized in the review \cite{senovilla}, in addition to energy conditions the singularity theorems assume initial formation of a trapped surface, in order for the geometry to inevitably evolve to a singularity.  In our model there is no trapped surface, so the formation of a singularity is avoided, even though the null and strong energy conditions are obeyed.}

\section{TOV Equation System in Spherical Polar Coordinates}

\subsection{Coordinate Fixing}

Calculations in general relativity begin by imposing coordinate conditions, to eliminate the ambiguities associated with covariance of the equations of relativity under general coordinate transformations. In calculations of stellar structure, the convenient choice of coordinates is spherical polar coordinates, in which the angular part of the line element has the form
\begin{equation}\label{ang}
ds^2=-r^2 (d\theta^2+\sin^2\theta d\phi^2)+..,
\end{equation}
with the ellipsis denoting radial and time terms.  This choice of coordinates is used in both the texts on relativistic stars of Zeldovich and Novikov \cite{zeld} and of Camenzind \cite{camen}, and is the one they use to derive the TOV \cite{oppen} equations.  Once the coefficient of $d\theta^2+\sin^2\theta d\phi^2$ is fixed as $-r^2$, no further coordinate transformation ambiguity is allowed, and the calculational results have a direct physical significance, unambiguously describing physics in the chosen coordinate system.

\subsection{The General Form TOV Equations}

The Tolman-Oppenheimer-Volkoff equations \cite{oppen}, \cite{zeld}, \cite{camen} are the relativistic generalization of the  hydrostatic equilibrium equations used in nonrelativistic calculations \cite{wein} of stellar structure.  For a spherically symmetric fluid with pressure $p(r)$ and energy density $\rho(r)$, they take the  general form

\begin{align}\label{newTOV}
\frac{dm(r)}{d r}=&4\pi r^2\rho(r)~~~,\cr
\frac{dp(r)}{d r}=&-\frac{\rho(r)+p(r)}{2} \frac{d\nu(r)}{d r} ~~~,\cr
\frac{d\nu(r)}{d r}=&\frac{N(r)}{1-2m(r)/r}~~~,\cr
N(r)=&(2/r^2)\big(m(r)+4\pi r^3 p(r)\big)~~~.\cr
\end{align}
Here  $m(r)$ is the volume integrated energy density within radius $r$,   and $\nu(r)=\log\big(g_{00}(r)\big)$.   The  general form TOV equations become a closed system when  supplemented by an equation of state $\rho(p)$ giving the energy density $\rho$ in terms of the pressure $p$. In the model studied in
\cite{adler1} the equation of state used is a relativistic matter equation of state $\rho(p)=3p$ for $p\leq {\rm pjump}$, and $p + \rho(p) =\beta$, for $p > {\rm pjump}$, with $\beta$ a small positive constant that is a parameter of the model.  The constant $\beta$ is a stand-in for more complex equation of state physics, and its positive value maintains decrease of pressure from the center of the black hole mimicker to the exterior.

From these equations we see that the Gliner equation of state $\rho + p =0$  has an interpretation completely independent of the cosmological constant vacuum: From the viewpoint of the TOV equations, it is the limiting case of the pressure evolution equation in which the pressure does not evolve with radius $r$.

\subsection{Continuity Properties Implied by the TOV Equations}

The form of the TOV equations implies that the quantities $p(r)$, $m(r)$, and $\nu(r)$ appearing on the left hand side must be continuous functions of $r$,  provided that all physical quantities are bounded.   To demonstrate this with a uniform notation for the three equations,
consider the first order general  differential equation
\begin{equation}\label{sample1}
\frac{dF(r)}{dr}=G(r)~~~,
\end{equation}
on the domain $r_A \leq  r \leq r_B$.  Then if $|G(r)|$ is bounded by ${\cal B}$ on this domain, the solution $F(r)$ must be continuous. To prove this, pick an arbitrary point $r_0$ in the domain,  integrate Eq. \eqref{sample1} from $r_0-\xi$ to $r_0+\xi$ with $\xi >0$, and take the absolute value, giving
\begin{equation}\label{sample2}
|F(r_0+\xi)-F(r_0-\xi)| \leq\int_{r_0-\xi}^{r_0+\xi} dr| G(r) |\leq 2 \xi {\cal B}~~~.
\end{equation}
Letting $\xi \to 0$, the right hand side of Eq. \eqref{sample2} vanishes, showing that $F(r)$ is continuous at $r_0$.

The fact that $p(r)$ must be continuous implies that the transition from a black hole mimicker exterior region to an interior one cannot involve a jump in the pressure.  This has a simple intuitive interpretation:  Pressure is the force per unit area, and and if this force is not continuous across any element of area, that area will accelerate, unless the area contains an infinite mass.  But this is ruled out by the assumption that all physical quantities are bounded.  Although we have just given a nonrelativistic argument, the conclusion reached is not altered by general relativity corrections, which are exactly incorporated into the TOV equations.

However, there is one quantity appearing in the TOV equations that is not restricted to be continuous by the structure of the equations.  This is the energy density $\rho(r)$, which only appears on the right hand side of the TOV equations.  So in black hole mimicker models constructed using the TOV equations, the energy density $\rho(r)$ can have discontinuous jumps, arising from a phase transition triggered by high enough pressure $p(r)$.  The TOV equations thus choose between the two possibilities for realizing the Gliner interior state described in Sec. 2, picking the case of positive pressure along with a transition to negative interior energy density.   As already pointed out,   that energy densities can be negative when quantum effects are included, and that more generally the dominant energy condition need not be satisfied in quantum theory,  has been known for many years \cite{ruderman}, \cite{wald}.

\subsection{Rescaling Covariance of the TOV Equations}

It is useful in numerical work to employ the following rescaling covariance of the TOV equations. Under  rescalings by a general parameter $L$,
\begin{align}\label{rescaling}
r \to &r/L~~~,\cr
m(r) \to & m(r/L)/ L~~~,\cr
\nu(r) \to & \nu(r/L)~~~,\cr
p(r) \to & p(r/L) L^2~~~,\cr
\rho(r) \to & \rho(r/L) L^2 ~~~.\cr
\end{align}
the TOV equation system of Eq. \eqref{newTOV} is form invariant.  This can be easily checked by direct substitution  into Eq. \eqref{newTOV}.   Rescaling covariance allows one to set the central pressure $p(0)$ to unity, as was done in the numerical calculations of \cite{adler1},  which amounts to specifying the size  of the unit in which distances $r$ are measured.

\subsection{The TOV Equations for a Relativistic Gas and  Their Rescaling Invariant Form}

In the latter parts of this review we will discuss in more detail the TOV equations for the exterior region of the model of \cite{adler1}, where $\rho(r)=3p(r)$. Making this substitution into Eq. \eqref{newTOV}, we get

\begin{align}\label{newTOV1}
\frac{dm(r)}{d r}=&12\pi r^2 p(r)~~~,\cr
\frac{dp(r)}{d r}=& - 2 p(r) \frac{d\nu(r)}{d r} ~~~,\cr
\frac{d\nu(r)}{d r}=&\frac{N(r)}{1-2m(r)/r}~~~,\cr
N(r)=&(2/r^2)\big(m(r)+4\pi r^3 p(r)\big)~~~.\cr
\end{align}

The coupled equations for $p(r)$ and $m(r)$ implied by Eq. \eqref{newTOV1} can be recast in a form manifestly invariant under the scale transformation of Eq. \eqref{rescaling}.  Define new  quantities
\begin{align}\label{invs}
t=&\log{r} ~~~,\cr
dt =& dr/r ~~~,\cr
\alpha(t)=& m(r)/r~~~,\cr
\delta(t)=&4 \pi r^2 p(r)~~~.\cr
\end{align}
Scale transformation corresponds to shifting the origin of $t$, and the remaining three quantities $dt$, $\alpha(t)$, and $\delta(t)$ are invariant under this shift.  In terms of these variables, the TOV equations for $p$ and $m$ take the scale-invariant form
\begin{align}\label{scaleinvTOV}
\frac{d\alpha(t)}{dt}=& 3 \delta(t)-\alpha(t)~~~,\cr
\frac{d\delta(t)}{dt}=& -4 \delta(t) \frac{\delta(t)+2 \alpha(t)-1/2}{1-2\alpha(t)}~~~.\cr
\end{align}
In Appendix B we give the Mathematica commands to compute a solution to these equations and plot the results.
The scale-invariant form of the TOV equation for $\nu$  is
\begin{equation}\label{nueq}
\frac{d\nu(t)}{dt}= 2 [\alpha(t)+\delta(t)]~~~,
\end{equation}
which can be integrated in terms of the solution of Eq. \eqref{scaleinvTOV}.

Remarkably,  these equations are what are called ``autonomous'' differential equations, in that no functions of the independent variable $t$ appear in the coefficients! This autonomous equation system describes a 2-dimensional flow,  a feature that may facilitate their  further mathematical analysis.   We note below that the
scale-invariant equations are easier for the Mathematica integrator to follow, allowing a survey of a wider domain of initial values without incurring diagnostics, than was permitted \cite{adler3} by the original form of the TOV equations.  They also have important implications for the smoothness of solutions to the equations, by showing that as long as $\alpha(t)$ avoids the value $\alpha=1/2$, where there is a pole in Eqs. \eqref{scaleinvTOV}, the solutions are analytic, that is continuous $C^\infty$ to all orders. For example, the sharp looking cusp shown in  Fig. 1, obtained from the program in Appendix B,  is $C^\infty$ despite the appearance of discontinuous behavior.   In the mimicker exterior,   this significantly  extends the conclusion reached in Sec. 3C that the quantities on the left hand side of the TOV equations must be continuous.

\begin{figure}[]
\begin{centering}
\includegraphics[natwidth=\textwidth,natheight=300,scale=0.8]{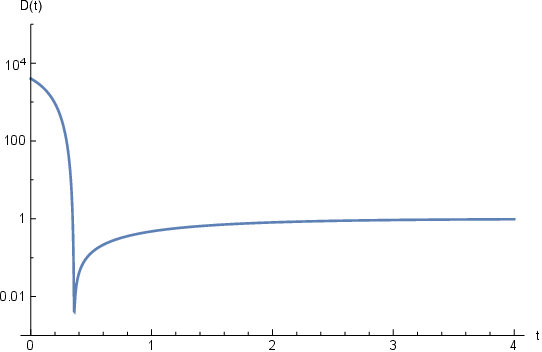}
\caption{Kink solution obtained from the scale-invariant TOV equations, with initial values $\alpha_0=-2,000, \, \delta_0=2,000$. A  deep and sharp cusp is obtained.}
\end{centering}
\end{figure}

\section{The ``Dynamical Gravastar'' Model}

\subsection{Equation of State and Boundary Conditions}

 The  ``dynamical gravastar'' black hole mimicker model introduced in \cite{adler1} is based, as noted above, on an equation of state with a jump in energy density as a function of pressure, according to
\begin{align}\label{eqnstate}
 \rho(p)=&3p, ~~~~ p\leq {\rm pjump}~~~,\cr
 \rho(p) =&-p+\beta, ~~~~  p > {\rm pjump}~~~. \cr
\end{align}
This equation of state was used \cite{adler1} to solve the TOV equations given above in Eq. \eqref{newTOV} with  results calculated for values of $\beta$ equal to $0.1$, $0.01$, and $0.001$.

The initial value conditions for the TOV equations can be taken as $ p(0)=1$ (which fixes the length scale used), $ m(0)=0$, and $\nu(0)=$``nuinit''.  The initialization nuinit is fixed {\it a posteriori} by requiring a match to the Schwarzschild metric value $\nu(r=\infty)={\rm log}\big(g_{00}(r=\infty)\big)=0$ at asymptotically large $r$, by a method that we now descibe.

\subsection{A simple two-step method for tuning $\nu(0)={\rm nuinit}$}

In our initial paper \cite{adler1} which included a cosmological constant, fixing nuinit required a multistep iteration, but in the simplified model of \cite{adler2} with no cosmological constant and with TOV equations taking the form of Eq. \eqref{newTOV}, a much simpler method is possible.

Inspecting  Eqs. \eqref{newTOV}, we see that $\nu(r)$ enters only through its derivative $d\nu(r)/dr$.   Hence a constant shift in $\nu(r)$, such as changing the initial value $\nu(0)$,  only affects $\nu(r)$ itself; the equations and numerical results for $m(r)$ and $p(r)$ are unaffected.  This permits a simple method for tuning the initial value $\nu(0)$ to achieve the large $r$ boundary condition $\nu({\rm rmax})=0$, where we have taken rmax as a proxy for $r=\infty$.

As seen in Fig. 9 of \cite{adler1}, when nuinit is correctly tuned, exterior to the mimicker nominal horizon at radius  $r=2M$ the metric coefficient $g_{00}=e^{\nu(r)}$ almost exactly coincides with the Schwarzschild metric value $g_{00}=1-2M/r$, with $M$ the gravastar mass.  This is a consequence of the fact that the density $\rho(r)$ becomes very small beyond $2M$, so by the Birkhoff uniqueness theorem for matter-free spherically symmetric solutions of the Einstein equations, the metric must take the Schwarzschild form.  A consequence of this, and of the TOV equation properties under shifts in nuinit, is that with an arbitrary initial guess for nuinit, the large $r$ value of $e^{\nu(r)}$ must have the form
\begin{equation}\label{largeR}
e^{\nu(r)}=e^K (1-2M/r)~~~,
\end{equation}
where $K$ is a constant, and where the gravastar mass $M$ can be read off from the exterior value of $m(r)$, since $m(r)$ is not affected by shifts in nuinit.  Thus, subtracting $K$ from the initial guess for nuinit,   with
\begin{equation}\label{Kvalue}
K=\nu(r)-\log(1-2M/r) \simeq \nu({\rm rmax})-\log\big(1-2 m({\rm rmax})/{\rm rmax}\big)~~~,
\end{equation}
gives the correct tuning of nuinit.  So the two step procedure to tune nuinit is: (i)  first run the TOV-solver program with an initial guess for nuinit, and calculate $K$ from Eq. \eqref{Kvalue}, and then (ii)   subtract this correction from the initial guess for nuinit and rerun the program.  We found that this method worked well in practice, with residual values of $K$ after the second step of order $10^{-6}$ or smaller.

\subsection{Some sample results for the ``dynamical gravastar'' mimicker model}

The Mathematica notebook commands for the simplified model are given in Appendix A.   In this notebook results of the computation are plotted for $D(r)=1-2m(r)/r$, for $g_{00}(r)= \exp(\nu(r))$, and for $\log\big(g_{00}(r)\big)=\nu(r)$. The plots are shown respectively in Figs. 2, 3, and 4, and a plot \cite{adler4} in which  panels copied from Figs. 2--4 are stacked vertically, with horizontal axes aligned, is given in Fig. 5.  This final plot shows that a ``simulated horizon'' or ``mock horizon'' appears at the radius $2M$, which lies well above the radius where the equation of state jumps.   Outside the mock horizon radius the metric is very close to that of a Schwarzschild solution, while inside the mock horizon  the behavior is very different, with $g_{00}$ becoming exponentially small but never negative.

The transition region around the mimicker simulated or mock horizon is the analog in our computation of the thin  ``skin'' of gravastars postulated in prior literature \cite{mazur}, \cite{camen}. It has appeared here in a natural way as a consequence of the equation of state jump defining the lower radius boundary of the TOV equations for relativistic matter.  Our results show  that the formation of a gravastar mock horizon is a  generic property of the exterior region TOV equations for relativistic matter, and
does not require the specific  interior region equation of state assumed in \cite{adler1}-- \cite{adler4}.     What is needed is for a sufficiently negative (or not too large a positive) volume integrated energy density $m(r)$ to be present at the inner boundary of the relativistic matter exterior region.  As already emphasized,  this is allowed  \cite{ruderman}, \cite{wald} because the energy density $\rho(r)$ can be negative when quantum corrections to the stress-energy tensor are taken into account.  The specific equation of state chosen {\it does} play an important role in determining the radius at which the mock horizon appears, and in determining how deep and sharp is the associated cusp or kink, since it determines the values of $m(r)$ and $p(r)$, or equivalently $\alpha(t)$ and $\delta(t)$, at the inner radius where the density jumps.

A parameter survey of the scale-invariant form of the TOV equations is given in \cite{adler3}, which found much better convergence properties than were obtained with the original form of these equations, and which suggests the following conjecture:

{\bf Conjecture:  Provided that the inner radius initial values $\alpha_0, \, \delta_0$ obey $\alpha_0<1/2$,   $\delta_0>0$, and $3\delta_0-\alpha_0>0$, the TOV equations for relativistic matter always yield a kink solution describing a  ``simulated'' or ``mock'' horizon.}

A  mathematical physics challenge is to prove this conjecture, and to obtain analytic results for the kink depth and width, as functions of the inner radius initial values $\alpha_0$ and $\delta_0$.

\begin{figure}[]
\begin{centering}
\includegraphics[natwidth=\textwidth,natheight=300,scale=0.8]{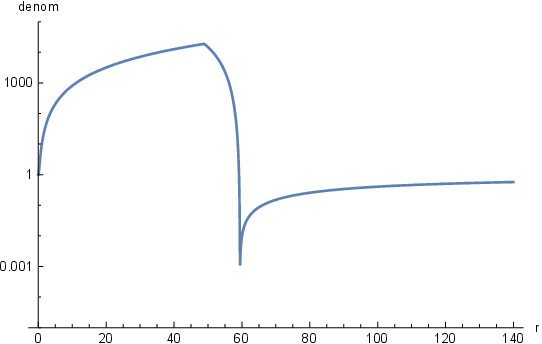}
\caption{Plot of $D(r)$ for the TOV.01 notebook.  The energy density jump at $r=48.895$ is apparent from the slope discontinuity in the upper region of the curve. This transition radius is not an input to the program, but rather emerges as a dynamical consequence of the jump in equation of state at pressure pjump. Above this radius the functions entering the TOV equations are continuous, as can be verified by zooming in on the apparent cusp near $r\simeq 59.43754$ with a finer plotting scale. For further discussion of smoothness of the exterior region solutions, where we show that they are $C^{\infty}$, see Sec. 3E.}
\end{centering}
\end{figure}

\begin{figure}[]
\begin{centering}
\includegraphics[natwidth=\textwidth,natheight=300,scale=0.8]{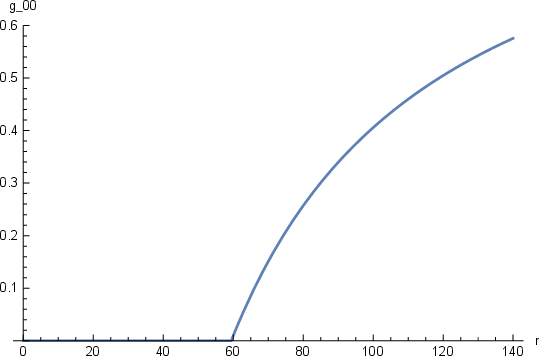}
\caption{Plot of $g_{00}$ for the same parameters used in Fig. 2.  Outside the radius $r\simeq 59.43754\equiv 2M$, this plot is nearly indistinguishable from a plot of the exterior Schwarzschild geometry for a black hole of mass $M=29.72$.   }
\end{centering}
\end{figure}

\begin{figure}[]
\begin{centering}
\includegraphics[natwidth=\textwidth,natheight=300,scale=0.8]{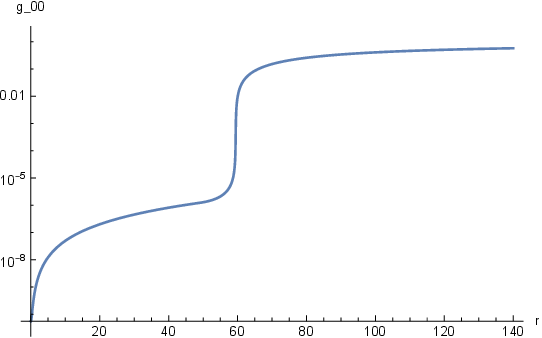}
\caption{Log plot of $g_{00}$ for the same parameters used in Fig. 2. This shows that $g_{00}$ is always positive within the mock horizon, but becomes very small.}
\end{centering}
\end{figure}

\vfill\eject
\vfill\eject

\section{ Concluding Remarks}

We conclude this mini-review with three remarks.

\begin{itemize}

 \item{\bf The horizon fantasy.}  If astrophysical black holes are not true black holes, but rather mimickers with no horizon, then the huge literature on mathematical black holes with horizons is fantasy.  Fantasy  can be great and enduring literature, but reality is more prosaic.

 \item{\bf The prosaic reality.}  As we have sketched in this mini-review, and in the more detailed articles on which it is based, the TOV equations for relativistic matter give a natural  mechanism for forming black hole mimickers with no horizon.  If there is a phase transition of matter at super-high pressure to a state with negative energy density, then the TOV equations for relativistic matter have  kink solutions that generate a ``simulated'' or ``mock'' horizon.  Outside the mock horizon, the metric  closely resembles that of a mathematical black hole, but inside the mock horizon the metric component $g_{00}$ remains positive while  becoming exponentially small down to the center of the mimicker.  The transition layer of the kink solutions can be very thin, with a structure completely determined by the TOV equations, without the artificial dialing of transition radii that is invoked in earlier black hole mimicker literature.  Interestingly, the exterior and interior equations of state in the dynamical gravastar model both constitute limiting cases: the relativistic matter equation of state $\rho=3p$ is the limiting  case of the widely applicable \cite{zeld1} matter equation of state inequality $\rho-3p          \geq 0$, and  the Gliner equation of state with  $p+\rho$  very small corresponds to the limiting case of the TOV equations where the pressure evolution versus radius becomes very small.

 \item{\bf Consequences for astrophysics.}  In this picture, astrophysical black holes can be ``leaky'' \cite{adlerleaky} with a  fraction of entering particles exiting in the form of a black hole ``wind''.   This allows the supermassive black holes that are believed to lurk at the center of every galaxy to play a direct role in nucleating galaxy formation, as suggested in articles by us \cite{adleressay}, \cite{adlerspace}  and by Silk et al. \cite{silk}.  It may also lead to long time delays \cite{cendes} between the infall of particles and phenomena induced by the exiting wind.  Observations of very early universe galaxy formation over the coming decade will play a decisive role in verifying, or falsifying this novel picture of the role of central astrophysical black holes in galaxy formation and evolution.

\end{itemize}

\section{Acknowledgements}

I wish to thank the following people for conversations and/or email correspondence in the course of the work reviewed here: Brent Doherty,  Ahmed ElBanna, Yuqi Li, Emil Mottola, Fethi Ramazano\u glu,  Kyle Singh, Thomas Spencer, James Stone, Scott Tremaine, Robert Wald,  Michael Weinstein, and George Wong.  During portions of the reviewed work I enjoyed the hospitality of the Aspen Center for Physics, supported by NSF grant No. PHY-1607611, and the hospitality of Clare Hall, Cambridge UK.

\appendix

\section{Mathematica notebook commands for Figs. 2, 3, 4}
The second line includes the value of nuinit arrived at by the procedure of Sec. 4B.  If parameters of the model (such as beta) are changed, the value of nuinit has to be readjusted.

\leftline{beta = .01;}
\leftline{nuinit = -23.62026189186466;}
\leftline{pjump = .95;}
\leftline{pinit = 1.;}
\leftline{rmax = 140.0;}
\leftline{rmin = .0000001;}
\leftline{alpha0 = -1;}
\leftline{alpha1 = 3;}
\leftline{kappa2 = 4*Pi;}

\medskip

\leftline{theta[x\_] := UnitStep[x];}
\leftline{alphas[x\_] := alpha0*theta[x - pjump] + alpha1*theta[pjump - x];}
\leftline{rho[x\_] := alphas[x]*x + beta*theta[x - pjump];}

\bigskip
system = \{

   \leftline{ nu[rmin] == nuinit,}

    \leftline{ p[rmin] == pinit,}

     \leftline{ em[rmin] == 0,}

   \leftline{em'[r] == kappa2 *r*r*rho[p[r]],}

 \leftline{ nu'[r] == (2/(r*r))*(em[r] + kappa2*r*r*r*p[r] )/(1 - 2 *em[r]/r),}

   \leftline{p'[r] == -(1/2)*(p[r] + rho[p[r]])*nu'[r]}

   \};

 \bigskip

\leftline{s = First@
   NDSolve[system, \{nu, p, em\}, \{r, rmin, rmax\}];}

    \medskip

\leftline{pout[r\_] := Evaluate[p[r] /. s];}
\leftline{nuout[r\_] := Evaluate[nu[r] /. s];}
\leftline{emout[r\_] := Evaluate[em[r] /. s];}
\leftline{denout[r\_] := (1 - 2 * emout[r]/r );}

\leftline{LogPlot[denout[r], \{r, rmin, rmax\}, PlotRange -$>$ \{.0001,100000\},
 AxesLabel -$  >$ \{"r", "denom"\}]}

\leftline{Plot[Exp[nuout[r]], \{r, rmin, rmax\}, PlotRange -$>$Automatic,
 AxesLabel -$>$ \{"r", "g\_00"\}]}

\leftline{LogPlot[Exp[nuout[r]], \{r, rmin, rmax\}, PlotRange -$>$ Automatic,
 AxesLabel -$>$\{"r", "g\_00"\}]}
\vfill\eject

\section{Mathematica notebook commands for Fig. 1}

\leftline{alpinit = -2000;}
\leftline{betinit = 2000;}
\leftline{tmin = 0;}
\leftline{tmax = 4;}
\medskip
system = \{

     \leftline{alp[tmin] == alpinit,}
     \leftline{bet[tmin] == betinit,}
   \leftline{Derivative[1][alp][t] == 3* bet[t] - alp[t],}
   \leftline{Derivative[1][bet][t] == -4*
     bet[t]*(bet[t] + 2 *  alp[t] - 1/2)/(1 - 2 *alp[t])}
   \};
 \medskip

\leftline{ s = First@NDSolve[system, \{alp, bet\}, \{t, tmin, tmax\}];}

   \bigskip
\leftline{   alpout[t\_] := Evaluate[alp[t] /. s];}
\leftline{denout[t\_] := 1 - 2 *alpout[t];}
\leftline{LogPlot[denout[t], \{t, tmin, tmax\}, PlotRange -$>$ \{.001, 50000\},
 AxesLabel -$>$ \{"t", "D(t)"\}]}

\vfill\eject
\bigskip

\vfill \eject
\bigskip

\section{\bf  Are ``Little Red Dots'' Dynamical Gravastars in Formation? (Added post-publication appendix)} \bigskip

{\bf We propose that the ``little red dots'' observed by the JWST are  dynamical gravastars  in formation. }

The aim of this short, qualitative, added appendix is to tie together two ideas that have appeared in the recent literature. The first, reviewed in \cite{adler11}, is the suggestion that horizonless black hole mimickers termed ``dynamical gravastars'' form when matter under extremely high pressure undergoes a phase transition to a state with {\it negative} energy density, as is permitted when quantum effects are taken into account.  The second is the suggestion \cite{lin}, \cite{leung}, \cite{beg} that the ``little red dots'' observed by the JWST are a new type of compact relativistic object, taking the form of an ultramassive  black hole surrounded by an opaque atmosphere.

The dyamical gravastar model of \cite{adler11} consists of solving the Tolman-Oppenheimer-Volkoff equations (which are a clever rewriting of the Einstein equations for a relativistic fluid source in a spherically symmetric metric) assuming an equation of state undergoing a phase transition at very high fluid pressure.  For pressure $p\leq {\rm pjump}$ the equation of state is taken as that of a relativistic fluid with energy density $\rho(p)=3p$, while for pressure $p> {\rm pjump}$ the energy density is taken as $\rho(p)= -p+\beta$, with $\beta$ a small  positive constant.  The model exhibits a metric structure very similar to that of a Schwarzschild black hole in an exterior region dictated by the equations, whereas  $g_{00}$ becomes exponentially small, but remains always positive, in the interior region. The calculations of \cite{adler1} are static, giving the equilibrium structure, but do not show what happens as this structure forms in a collapse.

Consider now the time-dependent dynamics of formation of a dynamical gravastar.  Once the pressure ${\rm pjump}$ is exceeded in a collapse, the model postulates a first order phase transition to a state of negative energy density.  There will be a large energy release in this collapse, leading to an exterior appearance of an extremely energetic object.  However, the fact that $g_{00}$ is exponentially small in the interior leads to a very large time dilation and red shift as the released energy leaks out.  Hence the energy release will be highly red shifted, and the time span of the energy release could be of a cosmological scale of millions or billions of years.  These characteristics can give an exterior appearance of a ``little red dot'' constituting an extremely  large energy source, consistent with what is observed.  It would be clearly of great interest, and well within current computer capabilities,  to do a computer simulation of this time-dependent scenario.

The papers of  \cite{lin}, \cite{leung}, and \cite{beg} are astrophysical analyses  motivated in part by the quasi-star model proposed earlier in \cite{begarm}.  The essence of our suggestion in this appendix is that the dynamical gravistar mechanism reviewed in \cite{adler1} gives a collapse dynamics that leads to a structure resembling a quasi-star, and can naturally accommodate the little-red-dot phenomenon currently being observed.

\newpage
\vfill{\hfill}
\begin{figure}[b]
\begin{centering}
 \includegraphics[natwidth=\textwidth,natheight=300,scale=0.7]{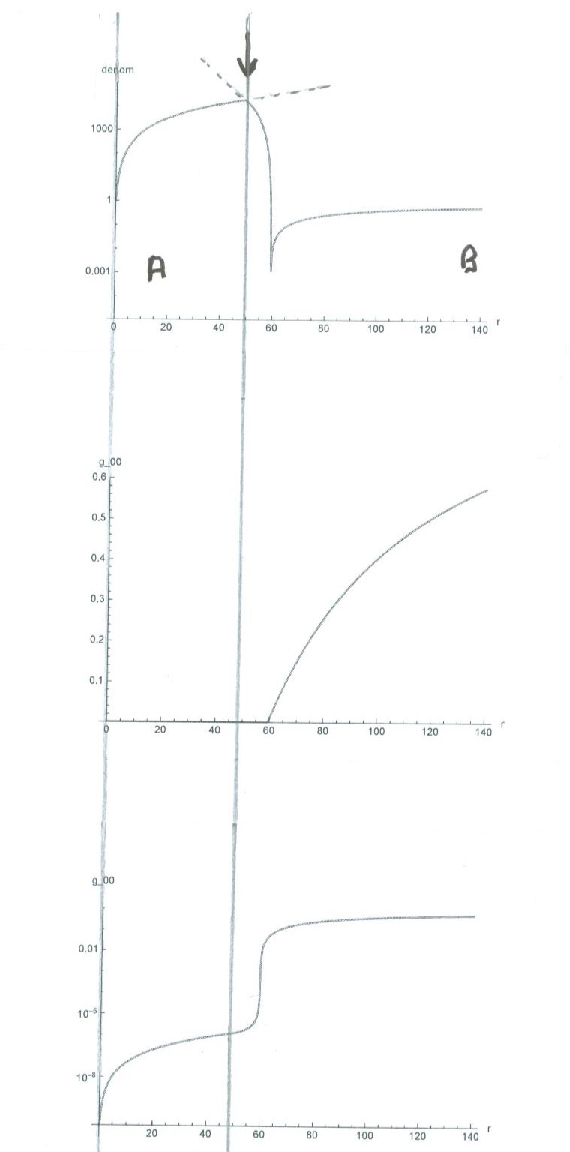}
\caption{In this figure the three plots of Figs. 2, 3, and 4 are stacked with their horizontal axes aligned.  The vertical line at $r=48.895$, marked by a down-pointing arrow, is where the discontinuity in equations of state at ${\rm pjump}=0.95$ appears; to the left of this line, in region A, the equation of state is $p+\rho=0.01$, and to the right of this line, in region B, the equation of state is $\rho=3p$.  The discontinuity in slopes of denom in the top panel shows up as the angle between the dashed lines being less than $\pi$, and arises because $m'(r)$ is not continuous where the equation of state is discontinuous.  The middle panel shows that at the cusp the metric component $g_{00}$ has nearly vanished on a linear plot, but the bottom panel shows that $g_{00}$ remains strictly positive down to zero radius, but precipitously drops to exponentially small values at radii below the radius $r\simeq 60$ of the cusp in denom. This gives an ``apparent horizon'', or perhaps better termed a ``mock horizon'', at the radius of the cusp.  Outside this radius the metric is very close to that of a Schwarzschild solution, while inside this radius the behavior is very different, with $g_{00}$ never going negative.}
\end{centering}
\end{figure}

\end{document}